\RequirePackage{fix-cm}
\documentclass{svjour3}                     
\smartqed  
\usepackage{graphicx}
\usepackage{natbib}
\usepackage[hyphens]{url}
\usepackage{hyperref}

\usepackage{amsmath}

\begin{document}

\title{A pragmatic Bayesian perspective on correlation analysis
}
\subtitle{The exoplanetary gravity - stellar activity case}


\author{P.~Figueira$^{1}$, J.~P.~Faria$^{1,2}$, V.~Zh.~Adibekyan$^{1}$, M.~Oshagh$^{1,3}$ \& N.~C.~Santos$^{1,2}$
}

\authorrunning{P. Figueira et al.} 

\institute{Pedro Figueira \at 
\email{pedro.figueira@astro.up.pt}   
  \and 
$^{1}$    Instituto de Astrof\'isica e Ci\^encias do Espa\c{c}o, Universidade do Porto, CAUP, Rua das Estrelas, 4150-762 Porto, Portugal \\
              $^{2}$    Departamento de F\'isica e Astronomia, Faculdade de Ci\^encias, Universidade do Porto, Rua do Campo Alegre, 4169-007 Porto, Portugal \\
              $^{3}$	Institut f\"ur Astrophysik, Georg-August-Universit\"at,  Friedrich-Hund-Platz 1, 37077 G\"ottingen, Germany 
}

\date{Received: date / Accepted: date}

\maketitle

\begin{abstract}
We apply the Bayesian framework to assess the presence of a correlation between two quantities. To do so, we estimate the probability distribution of the parameter of interest, $\rho$, characterizing the strength of the correlation. We provide an implementation of these ideas and concepts using {\tt python} programming language and the {\tt pyMC} module in a very short ($\sim$\,130 lines of code, heavily commented) and user-friendly program.

We used this tool to assess the presence and properties of the correlation between planetary surface gravity and stellar host activity level as measured by the log($R'_{\mathrm{HK}}$) indicator. The results of the Bayesian analysis are qualitatively similar to those obtained via p-value analysis, and support the presence of a correlation in the data. The results are more robust in their derivation and more informative, revealing interesting features such as asymmetric posterior distributions or markedly different credible intervals, and allowing for a deeper exploration. 

We encourage the reader interested in this kind of problem to apply our code to his/her own scientific problems. The full understanding of what the Bayesian framework is can only be gained through the insight that comes by handling priors, assessing the convergence of Monte Carlo runs, and a multitude of other practical problems. We hope to contribute so that Bayesian analysis becomes a tool in the toolkit of researchers, and they understand by experience its advantages and limitations. 

\end{abstract}

\section{Introduction}
\label{intro}

The search for extrasolar planets is a young field in astronomy, and yet its track record of results is remarkable. With roughly 2000 planets discovered as of today (September 2015), the most interesting properties started to emerge from the ensemble view. Several correlations surfaced from the extensive data gathered, often linking the properties of planetary bodies with those of their host stars. One of the most remarkable of these correlations is the one between planetary surface gravity and stellar activity as measured through the log(R'$_{HK}$) index. It was first established by \cite{2010ApJ...717L.138H} based on the data collected by \cite{2010ApJ...720.1569K} on 39 transiting planets. This correlation was recently revisited by \cite{2014A&A...572A..51F}, who extended the sample to one roughly 3 times larger, and performed a thorough analysis, including several consistency checks on the presence and interpretation of the correlation. The Spearman's correlation coefficient and associated p-value (obtained by performing 10\,000 Fisher-Yates shuffling of the data) showed that the correlation was present in the data at a significant statistical level. More attention was drawn to this issue by the contemporary work of \cite{2014A&A...572L...6L}, who proposed a physical interpretation of it; the author argued that the correlation found arises from the circumstellar material ejected by evaporating close-in planets. Because planets with lower surface gravity exhibit a greater mass-loss, the material adds up to a higher column density of circumstellar absorption, leading to a lower level of chromospheric emission as measured from our vantage point.

However, the analysis of \cite{2010ApJ...717L.138H} and \cite{2014A&A...572A..51F} suffer from the same intrinsic limitation: they use a p-value analysis to reject the null hypothesis of non-correlation in the data, and through this procedure infer if there is a correlation present. While being widely spread and by far the most used method in this kind of analysis, the use of p-values has been strongly criticized for its abusive usage \citep[e.g.][]{Raftery95bayesianmodel}. The recent editorial decision from the journal of {\it Basic and Applied Social Psychology} to ``ban Null Hypothesis significant testing procedure'' (i.e. p-values, t-values, etc.)\footnote{\url{http://www.tandfonline.com/doi/pdf/10.1080/01973533.2015.1012991}} should make uncomfortable the most stalwart defender of p-value analysis. In order to address this issue in a different and potentially more robust fashion, we take a different look at it using  Bayesian formalism.

In Section\,2 we describe the general Bayesian formalism and our implementation of the correlation analysis. In Section\,3 we describe the data used and report the values obtained by applying the framework of Section\,2. We conclude on Section\,4, and encourage the interested reader to play with the simple {\tt python} program we make available to the community, and explore how to assess the very common problem of the presence of a correlation in a dataset in a Bayesian way. 

\section{Pragmatic Bayesian Formalism}

The Bayesian definition of probability is strikingly different from the frequentist one. In the frequentist paradigm, a probability is the long-run frequency with which an event occurs (hence the name); in the Bayesian framework the probability of an event is a number that represents the degree of belief in the occurrence of that event, when all the available information about it is taken into account. This probability definition can be seen as subjective, but in its essence it is only conceptually different. The Bayesian framework allows one to incorporate the impact of different models describing our data, while explicitly including the prior knowledge and assumptions about the problem in a quantitative way. Very unfortunately, the profound conceptual differences between the two perspectives led to a polarization of scientific researchers into the seemingly adversary teams ``{\it frequentist vs. Bayesian}''; this antagonism did a huge disservice to the community, to which is far more interesting to weight the merits and shortcomings of each approach. Here we try to present a very short explanation of a practical application of the Bayesian machinery as a way of introducing it to a community that is not familiar with it, but is interested in an uncomplicated yet rigorous computational approach to the issue.

At the root of Bayesian inference is Bayes theorem. It can be derived from fundamental probability axioms, as demonstrated in \cite{Cox46}, and can be written in the form
\begin{equation}
P(A|D,I) = \frac{P(D|A,I)P(A|I)}{P(D|I)}
\end{equation}
in which the conditional probability $P(A|D,I)$ is called the {\it posterior distribution} and represents the probability of an event or hypothesis $A$, given the observed data $D$ and a model or set of assumptions encapsulated in our background information $I$. This posterior distribution, of interest to us, is equal to the product of $P(D|A,I)$, called the {\it likelihood}, times $P(A|I)$, called the {\it prior distribution}. The likelihood represents the probability of the data given the background information $I$, and given an event (or hypothesis) $A$; the prior $P(A|I)$ represents what we know about the event given the background information $I$. The term $P(D|I)$, often called the {\it evidence}, works as a normalization factor, ensuring that the posterior integrates to 1.0. In many cases, however, we are just interested in evaluating the shape of the posterior distribution, and since $P(D|I)$ does not depend on $A$, it can safely be ignored as a proportionality constant. This is exactly what happens in the particular case of parameter estimation. By dropping the evidence term we have that the probability distribution of a parameter $\rho$ -- that we now call {\it posterior distribution} in Bayesian parlay -- is given by
\begin{equation}
P(\rho|D,I) \propto P(D|\rho, I)P(\rho|I)
\end{equation}
in which the right-hand side of Eq.\,2 is given simply by the product of the probability of the data given the $\rho$ and background information -- the likelihood -- and of our specific information on the allowed and expected values of $\rho$ -- the prior. This presentation of the problem highlights that we are dealing with {\it probability distributions}, which can in principle be described analytically, but not necessarily so. Our final objective is to reach the {\it posterior distribution}, which can then be characterized in (the familiar) terms of expected values, scatter, and percentiles, just to mention some common statistics. We stress that those statistics are the result of the digestion of the posterior, which in itself contains all the information that can be obtained from our data. 
Our theoretical digression on Bayesian principles ends here, and we refer the interested reader to the remarkable book {\it Data Analysis: A Bayesian Tutorial}, by D.S. Sivia.
  
A significant practical problem blocks our way when we try to apply these concepts: the evaluation of the terms in Eq.\,1 or 2 might be impossible to perform analytically, and can even be unpractical to perform in a numerical way. As the number of parameters included in a model increases, the computational evaluation of the equations required to reach the posterior distribution becomes often intractable. So in order to do Bayesian analysis, we have to  address the practical problem of how to estimate our posterior distribution. At this point several options exist, but arguably the most common one is to perform a Markov Chain Monte Carlo (MCMC) that allows us to estimate the posterior distribution by drawing a large number of sample points from the right-hand-side of Eqs.\,1 or 2. As Cameron Davidson-Pilon writes in his {\it Probabilistic programming and Bayesian Methods for Hackers}\footnote{Freely available with examples and interactive code at \url{http://nbviewer.ipython.org/github/CamDavidsonPilon/Probabilistic-Programming-and-Bayesian-Methods-for-Hackers/tree/master/}.}: ``We should explore the deformed posterior space generated by our prior surface and observed data to find the posterior mountain. (...) MCMC returns samples from the posterior distribution, not the distribution itself. Stretching our mountainous analogy to its limit, MCMC performs a task similar to repeatedly asking {\it How likely is this pebble I found to be from the mountain I am searching for?}, and completes its task by returning thousands of accepted pebbles in hopes of reconstructing the original mountain.''

\subsection{Implementation in {\it python}}\label{sec:python}

To implement the previously described Bayesian approach to the evaluation of the presence of a correlation in a dataset we chose the open-source computer language {\tt python}. We used the well-known modules {\tt numpy}, {\tt Scipy}, and {\tt matplotlib} for basic data manipulation and plotting, and in order to sample our posterior distribution we chose the MCMC algorithm {\tt PyMC} \footnote{Installation and usage instructions can e found at \url{http://pymcmc.readthedocs.org/en/latest/index.html}. For more advanced usage and very insightful examples we refer to the previously mentioned Cameron Davidson-Pilon site.}.

For the data input we consider two float-value vectors $X$ and $Y$ with the same length. Within the program, each vector is standardized (i.e. to each value $X_i$ we subtract the average value $\overline{X}$ and divide this difference by the standard deviation $\sigma_X$); the frequentist statistics of Pearson's correlation coefficient and the Spearman's rank value calculated, as well as their associated p-values as delivered by the {\tt scipy} package. If chosen by the user, the standardization and ensuing operations are performed on the ranked variables.

We will assume that the variables $X$ and $Y$ are normally distributed, a valid assumption for a wide range of non-pathological scenarios. As such, we use the {\tt PyMC} distributions to represent our data (ranked or not) as generated by a bi-variate Gaussian distribution. Our distribution is characterized by the mean vector ($\mu_X, \mu_Y$) and its scale is represented by the covariance matrix
\[
CoV=
  \begin{bmatrix}
    \, \sigma_X^2 & \rho \, \sigma_X \sigma_Y \\
    \, \rho \, \sigma_X \sigma_Y & \sigma_Y^2 
  \end{bmatrix}
\]
Our objective is then to apply the formalism of the beginning of the chapter to calculate the posterior distribution $P(\rho|D, I)$ for the correlation coefficient (Pearson's correlation coefficient for non-ranked data, or Spearman's rank for ranked data, in both cases under the Bayesian framework). Using {\tt PyMC}, the likelihood evaluation is done by the MCMC algorithm given the priors for the variables of the bi-variate model: $\mu_X$, $\mu_Y$, $\sigma_1$, $\sigma_2$, and $\rho$. It is important to note that while doing so we defined the prior distribution for our variable of interest, $P(\rho|I)$. Since the data are standardized, we chose as prior for the mean values a Gaussian distribution centered at 0 with a dispersion of 1, and for the standard deviation a Inverse Gamma distribution with $\alpha$\,=\,11.0 and $\beta$\,=\,10.0. This is a strictly positive distribution, has an expected value of 1.0, and is a typical choice for an uninformative prior for scale parameters. For the $\rho$ parameter we chose a flat (uniform) distribution in the interval [-1,\,1]. It is important to note that we chose to work with uninformative priors in which the possible parameter values are only bounded by the definition of the corresponding quantities (such as the $\sigma$ being strictly positive and $\rho\,\in$ [-1,\,1]). As such these are expected to be applicable in a very wide range of scenarios. Once defined the model and its parameter's priors, we can feed our (X,Y) dataset to the MCMC sampler of {\tt pyMC}, and obtain the posterior distribution of each parameter, among which $\rho$.

Very interestingly, our problem is also tractable in a fully analytical way, a rather uncommon situation. For a bi-variate Gaussian distribution for $X$ and $Y$, the Inverse-Wishart distribution is a {\it conjugate prior}, i.e. if we define our priors in the form of a Inverse-Wishart distribution, the posterior distribution will also have a fully analytical form. As described in \citep{BergerSun}, if for the $\rho$ and $\sigma$ we choose a joint prior distribution given by
\begin{equation}
P(\mu_1, \mu_2, \sigma_1, \sigma_2, \rho|I) = \pi_{ab} = \frac{1}{\sigma_1^{3-a} \sigma_2^{2-b} (1-\rho^2)^{2-b/2}}
\end{equation}
we will have a constructive posterior distribution for $\rho$ given by
\begin{equation}
P(\rho |D,I) = Y^*/\sqrt{1+Y^{*2}}, \mathrm{ \:with \: } Y^* = - \frac{Z^*}{\sqrt{\chi^{2*}_{n-a}}} + \frac{\sqrt{\chi^{2*}_{n-b}}}{\sqrt{\chi^{2*}_{n-a}}}\frac{\rho_{D}}{\sqrt{1-\rho_D^2}}
\end{equation}
in which $\rho_D$ is the correlation coefficient of the data, and $n$ the number of data pairs. $Z^*$ will be independent draws from the standard normal distribution, and $\chi^{2*}_{n-a}$ and $\chi^{2*}_{n-b}$ will be independent draws from the chi-squared distributions with the indicated degrees of freedom. 
For ($a$=1, $b$=2) the prior distribution is called the {\it right-Haar prior}, and for ($a$=1, $b$=4) we have an uniform prior on $\rho$. As demonstrated in \citep{BergerSun}, the {\it right-Haar prior} is an objective prior (i.e., a prior that should be used when no additional information is available about the problem) that leads to an exact frequentist matching. These are two highly desired properties, and as such we use this prior as our default, but encourage the reader to experiment with the ($a$, $b$) values and compare the results. The reader is referred to \citep{BergerSun} and references therein for an in-depth discussion of the analytical solution, and to \cite{Ghosh} for a discussion on objective priors. Such priors should be handled with care, and while we provide this analytical result, we consider that the last word belongs to our numerical simulations. The program computes distributions as delivered by both methods as well as the most important statistics for comparison.

These operations are encapsulated in the program {\it BayesCorr.py}. This small program (only 130 lines of code) was written for readability and is extensively commented. We encourage the interested reader to tinker with it: test different prior distributions for the analysis, and explore MCMC parameters (the number of iterations, the burning phase limit, and the final thinning applied to the data). It was written to be useful even for those without a {\tt python} background, and is expected to deliver robust results without any need of fine-tuning. It is run simply by typing

\vspace{0.3cm}
\hspace{1cm} {\it python} BayesCorr.py
\vspace{0.3cm}

which performs a test run of the program using the datafile {\it testdata\_male.txt} from the example published in Rasmus B\aa \aa th blog\footnote{\url{http://sumsar.net/blog/2014/03/bayesian-first-aid-pearson-correlation-test/}} and that greatly inspired the code. To perform the analysis on user-provided data, one can type

\vspace{0.3cm}
\hspace{1cm} {\it python} BayesCorr.py {\it filename}
\vspace{0.3cm}

in which {\it filename} is a 2-column tab-formatted ASCII file listing the $X$ and $Y$ pairs. As runtime options, one can add a {\it r} string after the filename if the analysis is to be performed on ranked data, an {\it s} string if the posterior is to be saved in a one-column file, and an {\it rs} string if both functionalities are required. Once the program is terminated, the output of the analysis is presented on the shell itself for all the parameters, and their distributions plotted on the screen and saved as {\it .pdf} files. The main statistics (mean, standard deviation, 95\% Highest Posterior Density Credible Interval and quantiles) on the parameter $\rho$ are also saved in the comma-separated file {\it rho\_summary.csv}. By analyzing these results one can easily conclude if the data supports a correlation or not, and establish a link between a given correlation value and its expected probability.
The code is freely available for download or cloning in the git repository \url{https://bitbucket.org/pedrofigueira/bayesiancorrelation/}.

It is important to note that evaluating the posterior for $\rho$ is one way to assess the presence of a correlation and study its properties. An alternative is to consider Bayesian model comparison, in which the built-in Occam's razor would penalize more complex models -- in our case the correlated model, which has an extra parameter ($\rho$) when compared the non-correlated one. Since the case $\rho$\,=\,0 is a particular case of our model, we consider that all the information required for our purpose is present in the posterior distribution, which also informs us on the properties of the correlation. An analysis of the credible intervals for $\rho$ allows us to evaluate the probability of there being no correlation, $\rho$\,=\,0. This is a statement about the probability of a specific value of the parameter, which could not be made in the frequentist network. But we leave here our word of caution to the interested reader on how different approaches can be pursued to address the problem discussed.

\section{Planetary surface gravity vs. Stellar activity}

It is time to turn the machinery described to answer the problem that motivated the digression: the correlation between an exoplanet's surface gravity and its host star activity level. We perform the correlation tests considering two data sets -- the original \cite{2010ApJ...717L.138H} dataset, and the extended dataset of \cite{2014A&A...572A..51F}. For each of these we select 3 subsets respecting the following conditions: 

\begin{enumerate}
 \item massive, close-in planets ($M_\mathrm{p}$\,$>$\,0.1\,$M_\mathrm{J}$, $a$\,$<$\,0.1AU) orbiting stars with an effective temperature within the calibration range for thelog($R'_{\mathrm{HK}}$) indicator (4200\,K\,$<$\,$T_{\mathrm{eff}}$\,$<$\,6200\,K);
 \item planets orbiting stars with an effective temperature within the mentioned range (4200\,K\,$<$\,$T_{\mathrm{eff}}$\,$<$\,6200\,K);
 \item no restrictions.
\end{enumerate}

As discussed in \cite{2014A&A...572A..51F}, for the stars used in these study, different $T_{\mathrm{eff}}$ values were published, and this has an impact on the stars selected under Conditions 1) and 2); as shown in the same paper, SWEET-Cat \citep{2013A&A...556A.150S} temperatures are expected to provide most precise and accurate values due to an homogeneous analysis, and we consider these for the following analysis. For more details we refer to the mentioned paper. We fed the planetary surface gravity and activity index as variables $X$,$Y$ to the program, and performed the analysis of the ranked values, calculating the Spearman's correlation coefficient, as done in the previously mentioned works. 

\begin{figure*}

\includegraphics[width=\textwidth]{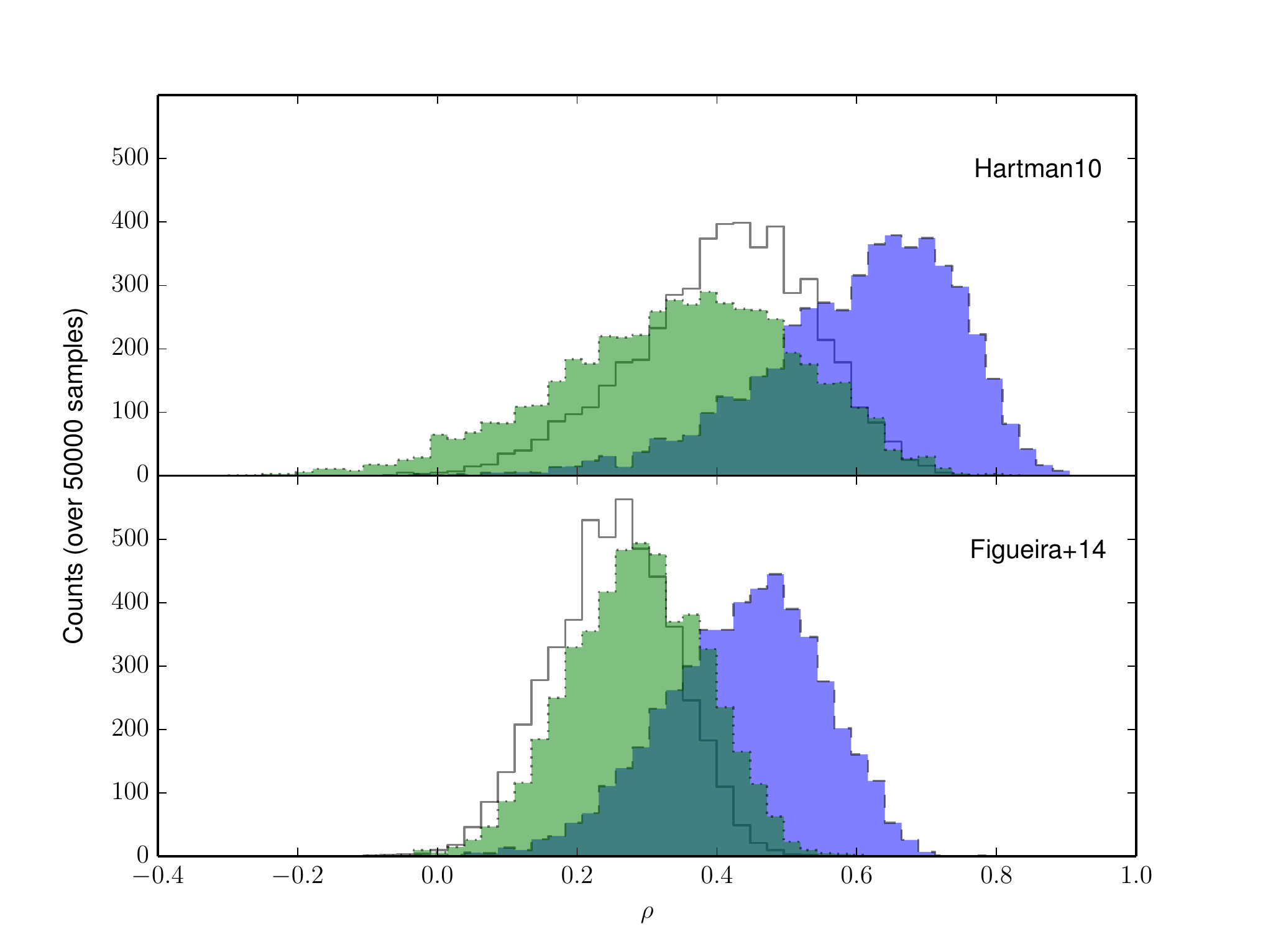}
\caption{Histogram of the posterior distribution of $\rho$ parameter for the datasets used in \cite{2010ApJ...717L.138H} ({\it upper panel}) and in \cite{2014A&A...572A..51F} ({\it lower panel}), when subject to the three conditions described in the text: 1) in blue / dashed line, 2) in green / dotted line, and 3) in white / solid line.}
\label{Fig:1}     

\end{figure*}

The posterior distributions are plotted in Fig.\,\ref{Fig:1}; their average value, standard deviation and 95\% credible intervals are presented in Table\,\ref{Tab:1}. We note that a credible interval X\% is the interval of the posterior probability in which the value of the parameter is comprised with a probability of X\%. In the same table we list, for comparison, the Spearman's rank correlation coefficient, z-score, and p-value obtained following the procedure described in \cite{2014A&A...572A..51F}.

\begin{table}
\caption{Statistical summary of the Posterior distribution ({\it left}) and comparison with \cite{2014A&A...572A..51F} results ({\it right}).}
\label{Tab:1}       
\begin{tabular}{c|ccc|ccc}
\hline\noalign{\smallskip}
dataset + Cond. & Mean & St. Dev. & 95\% CI  & Spearman's r. & z-score & p-value \\
\noalign{\smallskip}\hline\noalign{\smallskip}
Har2010 + 1 & 0.607 & 0.137 & [ 0.337, 0.842]  & 0.69 & 2.92 & 0.18 \\
Har2010 + 2 & 0.345 & 0.172 & [-0.011, 0.64]  & 0.41 & 1.86 & 3.14 \\
Har2010 + 3 & 0.409 & 0.127 & [ 0.157, 0.647]  & 0.45 & 2.76 & 0.28 \\
Fig+2014 + 1 & 0.435 & 0.113 & [0.216, 0.642]  & 0.47 & 3.25 & 0.06 \\
Fig+2014 + 2 &  0.286 & 0.097 & [0.091, 0.464] & 0.30 & 2.68 & 0.37 \\
Fig+2014 + 3 & 0.251 &0.088 & [0.08, 0.415] & 0.26 & 2.70 & 0.35 \\

\noalign{\smallskip}\hline
\end{tabular}
\end{table}

The results are very interesting. First of all, some of the posterior distributions are asymmetric: Condition 1) distributions are markedly so, and the other conditions for \cite{2010ApJ...717L.138H} dataset as well, but in a less pronounced way. This asymmetry is undetectable using frequentist analysis, and yet it is potentially insightful. It can suggest that the data does not follow a normal distribution, or that there are outliers in the data creating a distribution with pronounced tails. It certainly prompts the user to a closer analysis of the dataset. It is also clear that distributions associated to Condition 1 lean towards larger $\rho$ values, as delivered from frequentist analysis too. But the comparison between the two approaches in terms of results is not obvious at all: a larger Spearman's rank corresponds to a larger mean value of the distribution, but only in broad terms. Interestingly, the average value of the posterior is always smaller than the frequentist Spearman's rank value. The only case in which the $\rho$\,=\,0 case is inside the 95\% credible interval is also the one for which p-value is larger: Condition 2 applied to \cite{2010ApJ...717L.138H}. Other than that, the different datasets show that a correlation is very probable, with the 95\% credible interval lower limit being above $\rho$\,=\,0. It is also interesting to note that, as described in \cite{2014A&A...572A..51F}, the extended dataset leads to lower correlation coefficients, but this can be explained as a chance event for low-number dataset. More interesting is to note that there is a reduction of the width of the posterior peak, as measured by the standard deviation of the distribution. If we naively consider each data pair to be independent, then the width of the distribution should decrease with $\sqrt{n}$; by comparing the number of points in  each dataset, we would expect a width reduction by a factor of 1.7, larger than the average 1.5 factor measured, and even this one is subject to strong fluctuations. This is certainly a too simplistic analysis, but it is important to note that is only made possible because the posterior distribution of $\rho$ is calculated. We will refrain from comparing the different results in detail -- after all, they start from different definitions of probability -- but end the chapter with the note that the Bayesian analysis of these datasets also supports the presence of correlation between two quantities in the data.

\section{Conclusions}

We applied the Bayesian framework to assess the presence of a correlation between two quantities. To do so we, estimate the distribution of the parameter of interest, $\rho$, characterizing the strength of the correlation. We provide an implementation of these concepts using {\tt python} programming language and the {\tt pyMC} module in a very short ($\sim$\,130 lines of code, heavily commented) and user-friendly program. It was programmed thinking in those unfamiliar with the language, and yet leaving enough room for the more experienced to explore and play with it.

We used this tool to assess the presence and properties of the correlation between planetary surface gravity and stellar host activity level as measured by the log($R'_{\mathrm{HK}}$) indicator. The results of the Bayesian analysis are qualitatively similar to those obtained via p-value analysis, and support the presence of a correlation in the data. Yet, it is not a stretch to say they are not only more robust in their derivation, but more informative, revealing interesting features such as asymmetric posterior distributions or markedly different credible intervals, and allowing for a deeper exploration. 

We encourage the reader interested in this kind of assessment to apply our code to his/her own scientific problems. The full understanding of what Bayesian framework is can only be gained through the insight that comes by handling priors, assessing the convergence of Monte Carlo runs, and a multitude of other practical problems. We hope to contribute that Bayesian analysis becomes a tool in the toolkit of researchers, and they understand by experience its advantages and limitations. 

\begin{acknowledgements}
This work was supported by Funda\c{c}\~ao para a Ci\^encia e a Tecnologia (FCT) through the research grant UID/FIS/04434/2013. PF and NCS acknowledge support by Funda\c{c}\~ao para a Ci\^encia e a Tecnologia (FCT) through Investigador FCT contracts of reference IF/01037/2013 and IF/00169/2012, respectively, and POPH/FSE (EC) by FEDER funding through the program ``Programa Operacional de Factores de Competitividade - COMPETE''. PF further acknowledges support from Funda\c{c}\~ao para a Ci\^encia e a Tecnologia (FCT) in the form of an exploratory project of reference IF/01037/2013CP1191/CT0001. JPF acknowledges the support from FCT through the grant reference SFRH/BD/93848/2013. VA acknowledges the support from the Funda\c{c}\~ao para a Ci\^encia e a Tecnologia, FCT (Portugal) in the form of the fellowship SFRH/BPD/70574/2010. MO acknowledges support from Centro de Astrof\'{i}sica da Universidade do Porto through the grant of reference of CAUP-15/2014-BDP, and acknowledges research funding from the Deutsche Forschungsgemeinschft (DFG , German Research Foundation) - OS 508/1-1. This work results within the collaboration of the COST Action TD 1308. We thank everyone who contributed to develloping the open-source {\it python} language and keeping it free. Finally, we thank the referees for the constructive comments provided.
\end{acknowledgements}

\bibliographystyle{spbasic}      
\bibliography{Mybibliog}   

\end{document}